\documentstyle[12pt]{article}
\newcommand{\beq}{\begin{equation}}
\newcommand{\eeq}{\end{equation}}
\begin{document}
\title
{QED Penguin Contributions To Isospin Splittings of Heavy-Light Quark Systems}

\author{L. S. Kisslinger \\
      Department of Physics, Carnegie Mellon University\\
\vspace{5mm}
      Pittsburgh, PA 15213\\
	T. Goldman\\
	Theoretical Division, Los Alamos National Laboratory\\
\vspace{5mm}
	Los Alamos, NM 87545 \\
	Z. Li\\
      Department of Physics, Carnegie Mellon University\\
\vspace{5mm}
      Pittsburgh, PA 15213}

\maketitle
\indent
\begin{abstract}
\baselineskip=24pt
Recent experiments show that the isospin-violating mass splitting of
the B mesons is very small, but the best fits with a QCD sum rule
analysis give a splitting of at least 1.0 MeV. The isospin-violating
mass splittings of the charmed mesons, on the other hand, are in
agreement with experiment. In this letter we show that the inclusion of
2$^{nd}$ kind QED penguin diagrams can account for this discrepancy
within the errors in the QCD sum rule method.

\vspace{3mm} 

\noindent PACS numbers: 12.38.Lg, 13.40.-f, 14.40.Jz
\end{abstract}
\newpage
\baselineskip=24pt
\subsection*{1. Introduction}

The development of a gauge-invariant method for QED corrections to
composite systems\cite{kl1} has led to a consistent treatment within
the QCD sum rule method of the three sources of isospin mass
splittings:  the current quark mass differences in the QCD Lagrangian,
the nonperturbative QCD isospin violations which arise from u-d flavor
dependence of vacuum condensates, and electromagnetic effects. The
application of the method of Ref.\cite{kl1} removed serious
difficulties in earlier work on the isospin splittings of heavy-light
mesons\cite{kl2}. In a recent application\cite{kl3} to the charm and
bottom pseodoscalar and vector mesons, satisfactory agreement with the
experimental mass splittings was obtained within the expected accuracy
of the method.  However, although the D and D$^*$ isospin mass
splittings were found consistent with experimental data, the
theoretical result for the B$^+$-B$^0$ mass difference was about -1.2
MeV, while the experimental value is 0.35 $\pm$ 0.29 MeV\cite{cleo}.
Since the nonperturbative QCD effects tend to be quite small for the B
mesons, this difference between theory and experiment seems to be
somewhat larger than expected.

Several years ago it was pointed out\cite{gms} that there is a novel
electromagnetic effect which modifies the quark-gluon vertex in a
manner analogous to the penguin mechanism which leads to a weak
correction to the quark-gluon vertex.  This vertex modification is
referred to in Ref.\cite{gms} as an ``electromagnetic penguin of the
second kind'', because the term ``electromagnetic penguin'' was
previously applied to the weak corrections to the quark-photon vertex.
From the analytic form of the vertex modification\cite{gms}, it is
evident the the effect become increasingly important with increasing
quark mass. We explore here the possibility that the neglect of this
(second) QED penguin mechanism is the source of the discrepancy in the
B isospin splitting found in Ref.\cite{kl3}.

The main results of the present paper are to point out that it is
straightforward to find the largest nonperturbative QCD/QED effect as
well as perturbative effects arising from this QED penguin mechanism.
We estimate the contribution of nonperturbative processes to the
isospin-violating mass splittings of the D and B mesons and find a
contribution of approximately 0.5 MeV to the pseudoscalar B-meson
isospin mass splitting, while the penguins give a negligible
contribution to the D mesons.  The perturbative contributions are
somewhat larger. The result is a 1.2-2.0 MeV mass splitting for the B
and about 0.1 MeV for the D pseudoscalar mesons. From this we conclude
that the QED penguins are of the correct magnitude to provide an
explanation for the earlier discrepancy.

\subsection*{2. Nonperturbative QCD/QED Penguin Isospin-Violating Mass 
Splitting}

Since the only isospin symmetry violating mechanism being considered in
the present work is the (second) QED penguin vertex, it is convenient
to take as the starting point of the QCD sum rules the correlators used
for the heavy-light quark masses [without isospin
violations]\cite{kl4}.  For the pseudoscalar current $J_5(x)= : \bar
q(x) i \gamma_5 Q(x):$ the pseudoscalar current correlator is
\begin{eqnarray}\label{1}
\Pi_5(p)=i\int d^4 x e^{ipx} \langle 0 | T(J_5(x)J^+_5(0))| 0\rangle, 
\end{eqnarray} 
and
\begin{eqnarray}\label{2}
\Pi_{\mu\nu}(p)&=&i\int d^4 x e^{ipx} \langle 0 | T(V_{\mu}(x)V_{\nu}(0))
| 0\rangle \\ \nonumber &= &(q_{\mu}q_{\nu}-g_{\mu\nu}q^2)\Pi^{(1)}(q^2)
+q_{\mu}q_{\nu}\Pi^{(0)}(q^2)
\end{eqnarray}
is the correlator for the vector current $V_{\mu}(x)= : \bar
q(x)\gamma_{\mu} Q(x):$.  In the QCD sum rule method the correlator
$\Pi (p^2)$ is evaluated in two ways: 1) it is calculated starting from
QCD using the operator product expansion, and 2) it is treated
phenomenologically by a dispersion relation.

In the QCD calculation, the QED penguin isospin violations are obtained
by including the quark-gluon vertex modification, depicted in Fig. 1,
\beq\label{3}
  g^{pen} = V_q g,
\eeq
at appropriate places in the QCD evaluation. In Eq. \ref{3}, g is the
quark-gluon coupling constant and V$_q$, derived in Ref.\cite{gms}, is
the penguin vertex modification.

Keeping terms up to dimension $D = 5$, the microscopic nonperturbative
QCD evaluation of $\Pi (p^2)$ can be written as
\begin{eqnarray}\label{4}
\Pi(p^2) &=&C_I I + C_3 \langle\bar q q\rangle 
+C_4 \langle\alpha_s G^2\rangle + C_5 \langle  \bar q(\sigma \cdot G) q\rangle  
\end{eqnarray}
where the coefficient $C_I$ comes from the short distance correlation
calculated using perturbative quark propagators, $\langle\bar q
q\rangle$ is the quark condensate, $\langle\alpha_s G^2\rangle $ is the
gluon condensate, $\langle  \bar q(\sigma \cdot G) q\rangle$ is the
mixed condensate  and $C_n$ are the Wilson coefficients.  $C_I$ is
calculated from the process shown in Fig. 2a plus other two-loop
diagrams.  Since the graphs associated with the heavy-quark penguin
vertices do not contribute to the mass splittings of interest in the
present work, $C_I$ can be obtained by a simple modification of
previous calculations\cite{gb}. Introducing the variable $\omega$
\beq\label{5}
\omega=\frac {q^2 - M^2_Q}{2M_Q} 
\eeq
one can show that
the coefficients $C_3$ and $C_4$ are of the form
\begin{eqnarray}\label{6}
C_3 &\sim& \frac{1}{\omega}, \\ \nonumber
C_4 &\sim& \frac{1}{\omega}. \\ \nonumber
\end{eqnarray}
Since we eliminate these D=3 and D=4 processes in our sum rules, as we
show below, we do not give the detailed results for C$_3$ and C$_4$.
The coefficient $C_5$ from the process of Fig. 2b is
\beq\label{7}
C_{5a}=  V_{Q} \frac{1}{8 M_Q \omega^2} 
\eeq
for the pseudoscalar currents, and vanishes for the vector current.
For the process of Fig. 2c the D=5 mixed condensate coefficient is
\beq\label{8}
C_{5b}= -V_{q} \frac {1}{16 \omega^3 } 
\eeq
for both the pseudoscalar and vector currents.

Defining the Borel transformation by
\beq\label{9}
B f(\omega)={lim\atop{n\to \infty \atop {\omega\to\infty \atop 
\omega/n=\omega_B=\mbox{fixed}}}} \left [ \frac {\omega^{n+1}}{n!} \left (
-\frac d{d\omega}\right )^n\right ]f(\omega),
\eeq
and taking a derivative with respect to $\omega_B$ one obtains for the
nonperturbative penguin graphs through D=5
\beq\label{10}
 \frac{d \Pi^{pen}_{np}}{d\omega_B} = -\frac{\langle  \bar q(\sigma \cdot G) 
q\rangle}{16 \omega_B^3}(V_q + \frac{2 V_Q \omega_B}{M_Q})
\eeq
for the pseudoscalar and
\beq\label{11} 
 \frac{d \Pi^{pen}_{np}}{d\omega_B} = -\frac{V_q\langle  \bar q(\sigma \cdot G) 
q\rangle}{16 \omega_B^3}
\eeq
for the vector current. The Borel transformation of the D=3 and D=4
contributions [see Eq.\ref{6}] are independent of $\omega_B$ and
therefore do not contribute. Since we are only considering isospin
mass splittings in the present paper, the heavy quark vertex
modification, V$_Q$, does not contribute, and we can use the zero-mass
vertex modification from Ref.\cite{gms} for V$_q$.  The vertex
modification for V$_q$ in the limit of characteristic momenta larger
than the $\Lambda_{QCD}$ is\cite{gms}
\beq\label{12}
 V_q = -\frac{\alpha}{4 \pi} e_q^2 (1 + \frac{\pi^2}{3} + lnb + lnb^2),
\eeq
with b=$\frac{\Lambda_{QCD}^2}{M_Q^2}$. (Note that $\Lambda_{QCD}$ is
simply the relevant hadronic scale independent of quark masses and is
not necessarily directly related to standard QCD quantities such as
$\Lambda_{\bar{MS}}$.)  The characteristic momentum of the heavy-light
quark systems has been taken as the mass of the heavy quark, M$_Q$.

After the Borel transformation [Eq. \ref{9}], the phenomenological forms
of the correlators for the pseudoscalar and vector currents are
\beq\label{14}
\Pi^{ps}_{\mbox{phys}}=f_p^2\frac {M_P^4}{2M^3_Q} e^{-\frac
 {\Lambda_p}{\omega_p}}
\eeq
and
\beq\label{15}
\Pi^{v}_{\mbox{phys}}=f_v^2\frac {M_v^2}{2M_Q} e^{-\frac
 {\Lambda_v}{\omega_v}},
\eeq
respectively, where $\Lambda=\frac {M^2-M^2_Q}{2M_Q}$, $M$ is the mass
of the meson states, and $\omega_p$ and $\omega_v$ are the pseudoscalar
and vector Borel parameters. Defining $\Pi^{(0)}$ as the correlator
without the penguin processes, our sum rule for the mass shift due to
the penguins, $\Delta$M, is obtained by taking d$(\Pi
-\Pi^{(0)})$/d$\omega_B$,
\begin{eqnarray}\label{16}
\frac{2 f_p^2 M_p^3\Delta M_p}{M_Q^3
\omega_p^2}e^{-\frac{\Lambda_p}{\omega_p}} 
(\Lambda_p +\frac{M_p^2}{4 M_Q}) &=&  \frac{d \Pi_p^{pen}}{d\omega_p}
\end{eqnarray}  
for the pseudoscalar and
\begin{eqnarray}\label{17}
\frac{2 f_v^2 M_v\Delta M_v}{M_Q \omega_v^2}e^{-\frac{\Lambda_v}{\omega_v}}
(\Lambda_v +\frac{M_v^2}{2 M_Q}) &=&  \frac{d \Pi_v^{pen}}{d\omega_v}
\end{eqnarray}
for the vector meson. The penguin expressions on the right-hand side of
Eqs. \ref{16} and \ref{17} consist of the sum of the D=5
nonperturbative terms given in Eqs. \ref{10} and \ref{11} and the
perturpative contributions given by
\beq\label{20}
 \frac{d \Pi_{P.T.}^{pen}}{d\omega_B}= -\frac{1}{\pi} \int_{M_Q^2}^{s_0 +
M_Q^2} ds e^{-\frac{\omega}{\omega_B}} \frac{\omega}{2 M_Q \omega_B^2} 
ImC_I(s),
\eeq  
where $\omega$ is given by Eq. \ref{5} with s=q$^2$, and
s$_0$ is the threshold parameter for the continuum.
The expression for the perturbative correlator, C$_I$, can be
obtained from the results given in Ref\cite{gb} and are found in
Ref\cite{kl2}. 
\begin{eqnarray}\label{21}
 ImC_I(s)  & = & \frac{\alpha_s V_q}{2 \pi^2} s(1-x)^2 [\frac{9}{4}
 +2 l(x) + ln(x)ln((1-x) \\ \nonumber
 & + & (\frac{5}{2} -x - \frac{1}{1-x})ln(x) - (\frac{5}{2} - x)ln(1-x)],
\end{eqnarray}
where x=M$_Q^2$/s and $l(x)$ is the Spence function\cite{gb}.

We use the results of the sum rule analysis without the
penguins\cite{kl4} to fix the parameters needed for our estimate of the
mass shifts due to the QED penguins. For the D meson: M$_D$ = 1.867,
M$_Q$ = 1.5, F$_D$ = 0.13 and $\omega_D$ = 0.7, all in GeV. For the B
meson: M$_B$ = 5.279, M$_Q$ = 5.0, f$_B$ = 0.095 and $\omega_B$ = 0.63.
Neglecting the perturbative contributions and keeping the
nonperturbative contributions up to D=5, we obtain our results for the
nonperturbative QED penquin mass shifts from Eqs.\ref{10},\ref{12},\ref{16},\ref{20}:
\begin{eqnarray}\label{18}
   [\Delta M(D^+)-\Delta M(D^0)]^{np} &=& 0.07 MeV \\ \nonumber
   [\Delta M(B^+)-\Delta M(B^0)]^{np} &=& 0.5 MeV.
\end{eqnarray}
The perturbative contributions are sensitive to the choice of the
threshold parameter, s$_0$. The values of the threshold parameter which
we use are taken from a study of the continuum in Ref\cite{kl4}. The
are 1.8-2.1 GeV$^2$ for the D and 5.0-6.3 GeV$^2$ for the B systems.
We find for the perturbative contributions:
\begin{eqnarray}\label{19}
   [\Delta M(D^+)-\Delta M(D^0)]^{pt} &=& 0.1-0.2 MeV \\ \nonumber
   [\Delta M(B^+)-\Delta M(B^0)]^{pt} &=& 0.7-1.5 MeV 
\end{eqnarray}
for this range of the s$_0$ values.
Note that the largest nonperturbative QED penguin contributions are
negligible for the D isospin mass splittings, while they are of the
order of the discrepancy in Ref.\cite{kl3} for the B isospin mass
splittings. The net B mass splitting resulting from combining the
nonpenguin result, M(B$^+$)-M(B$^0$)=-1.2 MeV from Ref.\cite{kl3}, with
the penguin results of Eqs.\ref{18} and \ref{19} is
\begin{eqnarray}\label{22}
   [\Delta M(B^+)-\Delta M(B^0)] &=& 0.0-0.8 MeV, 
\end{eqnarray}
in agreement with experiment\cite{cleo}

\subsection*{3. Conclusions}

In the present analysis we have shown that the QED penguins are of the
proper magnitude to account for the experimental result that there is
essentially no isospin mass splitting in the pseudoscalar B meson
systems. From Eq.\ref{12}, it is evident that the QED penquin
contributions are much larger for the B than for the D systems, and
that the good results of Ref.\cite{kl3} for the D system are not
modified.  We conclude that the QCD sum rule method can account for the
heavy-light meson isospin mass splitting with the parameters that have
been used for light quark systems.

This work is supported in part by National Science Foundation grant
PHY-9319641 and in part by the Department of Energy.

\vspace{5mm}
\noindent {\Large\bf Figure Captions}

\vspace{5mm}
   1.  The QED penguin vertex. The oscillating curve represents 
a photon and the corkscrew curve represents a gluon. The straight 
line represents a quark. 

\vspace{5mm}

   2.  a) The perturbative QED penguin diagram;
       b), c) the D=5 nonperturbative penguin diagrams.
The heavy line represents the massive quark ($c$ or $b$) and 
the light line represents the light quark ($u$ or $d$). 

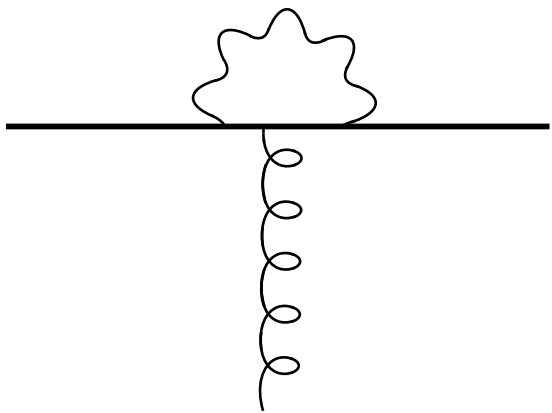
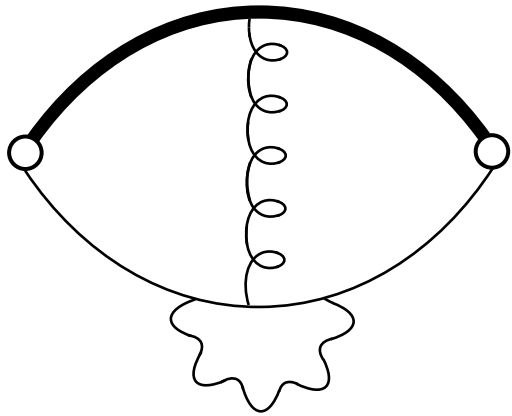
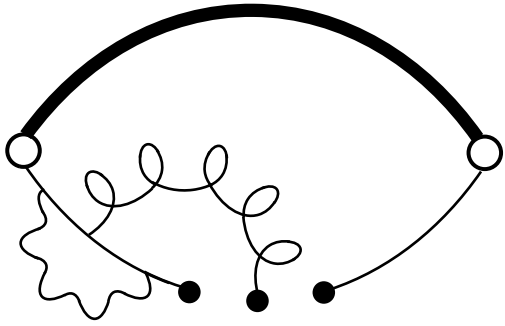
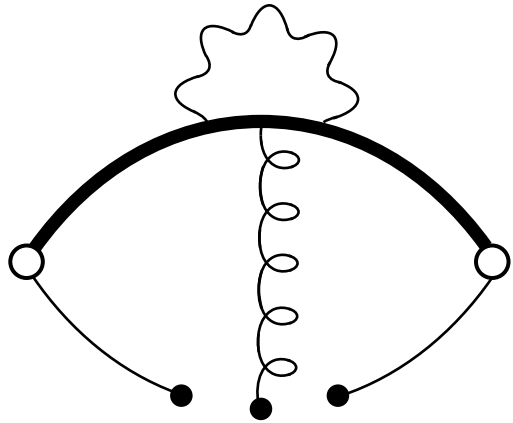
\end{document}